\documentclass[12pt]{article}
\usepackage[english,german,french,polish]{babel}
\usepackage[T1]{fontenc}
\usepackage{amsfonts}

\begin{document}

\selectlanguage{english}

\baselineskip 0.73cm
\topmargin -0.4in
\oddsidemargin -0.1in

\let\ni=\noindent

\renewcommand{\thefootnote}{\fnsymbol{footnote}}

\newcommand{\SM}{Standard Model }

\newcommand{\SMo}{Standard-Model }

\pagestyle {plain}

\setcounter{page}{1}



~~~~~~
\pagestyle{empty}

\begin{flushright}
IFT-- 09/14
\end{flushright}

\vspace{0.4cm}

{\large\centerline{\bf Nongauge antisymmetric-tensor bosons}}

{\large\centerline{\bf mediating interactions in the hidden sector}}

\vspace{0.5cm}

{\centerline {\sc Wojciech Kr\'{o}likowski}}

\vspace{0.3cm}

{\centerline {\it Institute of Theoretical Physics, University of Warsaw}}

{\centerline {\it Ho\.{z}a 69, 00--681 Warszawa, ~Poland}}

\vspace{0.6cm}

{\centerline{\bf Abstract}}

\vspace{0.2cm}

\begin{small}


In a recent work, we have constructed a model of hidden sector of the Universe, consisting of sterile spin-1/2
Dirac fermions ("\,$\!$sterinos"), sterile spin-0 bosons ("\,$\!$sterons") conjectured to get spontaneously nonzero 
vacuum expectation value, and also of conventional photons assumed to participate both in the  hidden and 
Standard-Model sectors (after the electroweak symmetry is spontaneously broken by the Standard-Model Higgs 
mechanism). This provides a "photonic portal"\, to the hidden sector as an alternative to the popular "Higgs portal".
Moreover, we have proposed to stop the proliferation of new kinds of gauge bosons as sterile mediators of nongravitational 
interactions in the hidden sector, introducing instead nongauge mediating bosons  described by an antisymmetric-tensor 
field (of dimension one), whose sources are given not only by sterino-antisterino pairs, but also by steron-photon pairs. 
Then, conventional photons display, beside the familiar gauge coupling to the Standard-Model electric current, a new 
nongauge weak coupling of sterile antisymmetric-tensor bosons to steron-photon pairs, where photons are described by 
the gauge-invariant electromagnetic field $F_{\mu\,\nu}$. The new mediators are unstable, decaying, for instance, into 
electron-positron pairs. When interacting in pairs, they can  annihilate simply into two photons. The corresponding cross-section is calculated.
 
\vspace{0.6cm}

\ni PACS numbers: 14.80.-j , 04.50.+h , 95.35.+d 

\vspace{0.6cm}

\ni December 2009

 
\end{small}

\vfill\eject

\pagestyle {plain}

\setcounter{page}{1}

\vspace{0.4cm}

\ni {\bf 1. Introduction}

\vspace{0.4cm} 

This note is a sequel of Ref. [1], where we have continued the discussion of a proposed model of hidden sector of the Universe, consisting of sterile spin-1/2 Dirac fermions ("\,$\!$sterinos"), sterile spin-0 bosons ("\,$\!$sterons"), and sterile nongauge mediating bosons ("$A$ bosons") described by an antisymmetric-tensor field (of dimension one) [2].

We have emphasized a two-level structure of the proposed model, where the hidden sector is weakly coupled to the \SMo sector through the "photonic portal"\, to the hidden sector, provided by the new nongauge weak interaction of the form

\begin{equation}
- \frac{1}{2} \sqrt{\!f\,}\left[(<\!\!\varphi\!\!>_{\rm vac}\! + \,\varphi_{\rm ph}) F_{\mu \nu} + \zeta \bar\psi \sigma_{\mu \nu} \psi \right] A^{\mu \nu}
\end{equation}

\ni with $\sqrt{\!f\,}$ and $\sqrt{\!f\,}\,\zeta$ denoting two dimensionless small (real) coupling constants. Here, $\psi$ and $\varphi = <\!\!\varphi\!\!>_{\rm vac}\! + \,\varphi_{\rm ph}$ with $<\!\!\varphi\!\!>_{\rm vac}\! \neq 0$ are sterino and steron fields, respectively, $A_{\mu \nu}$ stands for our sterile nongauge mediating field, while $F_{\mu \nu} = \partial_\mu A_\nu - \partial_\nu A_\mu $ is the \SMo gauge electromagnetic field (after the electroweak symmetry is spontaneously broken by the \SMo Higgs mechanism). In our case, the electromagnetic field interacts both in the hidden and \SMo sectors, participating in structures of both.

At the first level, the \SMo electric current proportional to $e = \sqrt{4\pi\alpha}$ acts as the source of the electromagnetic field $F_{\mu \nu}$ in the "\,$\!$supplemented Maxwell's equations"

\begin{equation}
\partial^\nu \left[F_{\mu \nu} +  \sqrt{\!f\,}(<\!\!\varphi\!\!>_{\rm vac}\! + \,\varphi_{\rm ph}) A_{\mu \nu}\right] = -j_\mu \;\;,\;\; F_{\mu \nu} = \partial_\mu A_\nu - \partial_\nu A_\mu 
\end{equation}

\ni related to the interaction (1). At the second level, the electromagnetic field $F_{\mu \nu}$ multiplied by the constant $\sqrt{\!f\,} <\!\!\varphi\!\!>_{\rm vac} \neq 0$ becomes, in a spontaneous way, a part of the source of the sterile nongauge field $A_{\mu \nu}$ mediating nongravitational interactions in the hidden sector. It is so, because the field equation 

\begin{equation}
(\Box - M^2)A_{\mu \nu} = - \sqrt{\!f\,}\left[ (<\!\!\varphi\!\!>_{\rm vac}\! + \,\varphi_{\rm ph}) F_{\mu \nu} + \zeta \bar\psi \sigma_{\mu \nu} \psi\right] 
\end{equation}

\ni  related to the interaction (1) holds for $A_{\mu \nu}$, where $M$ is a mass scale typically expected to be large. A bold, but attractive, option may be here $f = e^2 = 4\pi\alpha = 0.0917$ with $\alpha = 1/137$. Then, the mass scale $M$ ought to be large enough in order to diminish properly the effective form of weak interaction following from Eq. (1).

In fact, if momentum transfers mediated through the  field $A_{\mu \nu}$ in virtual states can be neglected {\it versus} the mass scale $M$, then it follows approximately from Eq. (3) that

\begin{equation}
 A_{\mu \nu} \simeq \frac{\sqrt{\!f\,}}{M^2}\left[ (<\!\!\varphi\!\!>_{\rm vac} + \varphi_{\rm ph}) F_{\mu \nu} + \zeta \bar\psi \sigma_{\mu \nu} \psi \right]\,.
\end{equation}

\ni In this case, the interaction (1) gives approximately the effective interaction (in an effective Lagrangian):

\begin{equation}
  -\frac{1}{4} \frac{f}{M^2}\left[(<\!\!\varphi\!\!>_{\rm vac}\! + \,\varphi_{\rm ph}) F_{\mu \nu} + \zeta \bar\psi \sigma_{\mu \nu} \psi \right] \left[(<\!\!\varphi\!\!>_{\rm vac}\! + \,\varphi_{\rm ph}) F^{\mu \nu} + \zeta \bar\psi \sigma^{\mu \nu} \psi \right] 
\end{equation}

\ni being an analogy to the familiar Fermi coupling. Here, $\varphi = <\!\!\varphi\!\!>_{\rm vac}\! + \,\varphi_{\rm ph}$ with $<\!\!\varphi\!\!>_{\rm vac} \neq 0$.

It can be seen from Eq. (5) that sterinos display the effective magnetic interaction

\begin{equation}
- \mu_\psi \bar\psi \sigma_{\mu \nu} \psi F^{\mu \nu}  
\end{equation}

\ni proportional to the small magnetic moment

\begin{equation}
\mu_\psi = \frac{f \zeta}{2M^2}<\!\!\varphi\!\!>_{\rm vac} 
\end{equation}

\ni generated spontaneously by $<\!\!\varphi\!\!>_{\rm vac} \neq 0$. Of course, sterinos (and sterons) are electrically neutral. This is consistent with the fact that the total source current for $F_{\mu \nu}$ in the "\,$\!$supplemented Maxwell's equations"~ (2) differs from the \SMo electric current $j_\mu$ only by the four-divergence $\sqrt{\!f\,} \partial^\nu\left[(<\!\!\varphi\!\!>_{\rm vac}\! + \,\varphi_{\rm ph}) A_{\mu \nu}\right]$ giving no contribution to the total electric charge. In contrast, if $A_{\mu \nu}$ were a gauge field mediating in the hidden sector and kinematically mixing with $F_{\mu \nu}$, the situation would be different [3].

The vacuum expectation value $<\!\!\varphi\!\!>_{\rm vac} \neq 0$ can also generate spontaneously the masses $m_\psi$, $m_\varphi$ and $M$ of sterinos, physical sterons and $A$ bosons [2]. 

Due to the "photonic portal", our approach to the hidden sector differs essentially from the popular approach, where "Higgs portal"\, to the hidden sector is introduced [4].

\vspace{0.4cm}  

{\bf 2. Nonrelativistic splitting of $A_{\mu \nu}$}

\vspace{0.4cm}

In analogy with the splitting of  the electromagnetic field $F_{\mu \nu} = \partial_\mu A_\nu - \partial_\nu A_\mu $ into its electric and magnetic three-dimensional fields $\vec{E} =\left(E_k\right)$ and $\vec{B} =\left(B_k\right)\;(k = 1,2,3)$,

\begin{equation} 
\left(F_{\mu \nu}\right) = \left(\begin{array}{rrrr} 0 & E_1 & E_2 & E_3 \\ -E_1 & 0\;\; & -B_3 & B_2 \\ -E_2 & B_3 & 0\;\; & -B_1 \\  -E_3 & -B_2 & B_1 & 0\;\; \end{array} \right) \,,
\end{equation}

\ni where 

\begin{equation} 
\vec{E} = -\partial_0 \vec{A} - \vec{\partial} A_0 = (-F_{k  0}) \;\;\;,\;\;\; \vec{B}  = \vec{\partial}\times \vec{A} = \left(- \frac{1}{2}\varepsilon_{k l m} F_{l m}\right)
\end{equation}

\ni with $\vec{A} = (A^k) = (-A_k)$ and $\vec{\partial} = (\partial_k) = (-\partial^k)\;(k=1,2,3)$, we can write 

\begin{equation} 
\left(A_{\mu \nu}\right) = \left(\begin{array}{rrrr} 0\;\;\;\;  & A^{(E)}_1 & A^{(E)}_2 & A^{(E)}_3 \\ -A^{(E)}_1 & 0\;\; \;\; & -A^{(B)}_3 & A^{(B)}_2 \\ -A^{(E)}_2 & A^{(B)}_3 & 0\;\; \;\; & -A^{(B)}_1 \\  -A^{(E)}_3 & -A^{(B)}_2 & A^{(B)}_1 & 0\;\; \;\; \end{array} \right) \,, 
\end{equation}

\ni where

\begin{equation}
\vec{A}^{(E)} = \left(- A_{k 0}\right) \;\;\; \;,\;\;\; \vec{A}^{(B)} = \left(- \frac{1}{2}\varepsilon_{k l m} A_{l m}\right)
\end{equation}

\ni $(k = 1,2,3)$. The vector and axial three-dimensional fields $\vec{A}^{(E)} = \left(A^{(E)} _{k}\right)$ and $\vec{A}^{(B)} = \left(A^{(B)} _{k}\right) \;\;(k=1,2,3)$ get spin 1 and, respectively, parity $-$ and + (if the interaction (1) or (13) preserves the parity). 

Then, making use of Eqs. (8) and (10), and also of the matrix

\begin{equation} 
(\sigma^{\mu \nu}) = \left(\begin{array}{rrrr} 0\;\;  &i \alpha_1 & i \alpha_2 & i \alpha_3 \\ -i \alpha_1 & 0\;\; & \sigma_3 & -\sigma_2 \\ -i \alpha_2 & -\sigma_3 & 0\;\; & \sigma_1 \\  -i \alpha_3 & \sigma_2 & -\sigma_1 & 0\;\; \end{array} \right) 
\end{equation}

\ni for spin tensor $\sigma^{\mu \nu} = (i/2)[\gamma^\mu,\gamma^\nu]$ with $\vec{\alpha} = (\alpha_k) = (\gamma^0\gamma^k) = (i\sigma^{k 0})$ and $\vec{\sigma}=(\sigma_k) = (\gamma_5\vec{\alpha}) = \left((1/2)\varepsilon_{k l m} \sigma^{l m}\right) \;(k=1,2,3)$, we can present the interaction (1) in the form

\begin{equation}
\sqrt{\!f\,}\left\{\left[(<\!\!\varphi\!\!>_{\rm vac}\! + \,\varphi_{\rm ph})  \vec{E}- i\zeta \bar\psi \,\vec{\alpha} \,\psi \right]\cdot \vec{A}^{(E)}- \left[(<\!\!\varphi\!\!>_{\rm vac}\! + \,\varphi_{\rm ph})  \vec{B} - \zeta \bar{\psi} \,\vec{\sigma}\psi \right]\!\cdot\!\vec{A}^{(B)}\right\} 
\end{equation}

\ni and, in consequence, the field equations (3) for $A_{\mu \nu}$ as

\begin{eqnarray} 
(\Box - M^2)\vec{A}^{(E)} & = & - \sqrt{\!f\,}\left[(<\!\!\varphi\!\!>_{\rm vac}\! + \,\varphi_{\rm ph})  \vec{E} - i \zeta \bar{\psi}\,\vec{\alpha}\,\psi \right]\,, \nonumber \\ 
(\Box - M^2)\vec{A}^{(B)} & = & - \sqrt{\!f\,}\left[(<\!\!\varphi\!\!>_{\rm vac}\! + \,\varphi_{\rm ph})  \vec{B}\, - \,\zeta\bar{\psi}\, \vec{\sigma}\,\psi \right]\,.
\end{eqnarray}

\ni Here, $\varphi = <\!\!\varphi\!\!>_{vac}\! + \,\varphi_{\rm ph}$ with $<\!\!\varphi_{\rm vac}\!\!> \neq 0$. 

When  the sterile $A$ bosons of two kinds described by the fields $\vec{A}^{\rm (E)}$ and $\vec{A}^{\rm (B)}$ propagate freely in space, they get the following wave functions:

\begin{equation} 
\vec{A}^{\,(E,B)}_{\vec{k}_A}(x) = \frac{1}{(2\pi)^{3/2}} \frac{1}{\sqrt{2 \omega_A}}\, \vec{e}^{\;(E,B)} e^{-i k_A\cdot x} \,,
\end{equation}

\ni where $k_A = (\omega_A, \vec{k}_A)$ with $\omega_A =\sqrt{\vec{k}_A^2 + M^2}$, while $\vec{e}^{\;(E,B)} = \vec{e}_a^{\;(E,B)}\;\;(a=1,2,3)$ are three orthonormal linear polarizations of $A^{(E)}$ and $A^{(B)}$ bosons,  satisfying the relations

\begin{equation}
\vec{e}_a^{\;(E,B)}\cdot \vec{e}_b^{\;(E,B)} = \delta_{a b}\;\; (a,b = 1,2,3) \;\;\;,\;\;\; \sum^3_{a=1} {e}_{a k}^{\,(E,B)} {e}_{a l}^{\,(E,B)} = \delta_{k l}\;\;(k,l = 1,2,3)
\end{equation}

\ni with $\vec{e}_a ^{\;(E,B)}\! = \!({e}_{a k}^{\;(E,B)}) \, (a=1,2,3)$. Here, $\!\vec{e}_a ^{\,(B)}\!$ ought to be axial vectors, if the parity is con\-served by the new weak interaction in hidden sector. 

Since such $\!\vec{e}_a ^{\,(B)}\!$ cannot be realized without imposing some (in fact, nonexisting) Maxwell-type relations between the massive fields $\vec{A}^{\,(E)}\!$ and $\vec{A}^{\,(B)}\!$ (of dimension one), the vectors $\vec{e}_a ^{\,(B)}\!$ are here polar. This implies that the coupling (13) of $\vec{A}^{\,(B)}\!$ {\it violates maximally} the parity in hidden sector (as now $A^{\,(B)}\!$ bosons cannot avoid to get parity $-\!$, like $\!A^{\,(E)}\!$ bosons, although both remain independent). 

\vspace{0.3cm}

\ni {\bf 3. Annihilation of $A$-boson pairs into double photons} 

\vspace{0.3cm}

Note that sterile mediating $A$ bosons are unstable. If $M > 2 m_f$, their simple decays are those into pairs of charged fermions,  $A \rightarrow \gamma^* \rightarrow \bar{f} f$ (where, for instance, $f = e^-$ or $p$), caused by the coupling

\begin{equation}
\sqrt{\!f\,}\,<\!\!\varphi\!\!>_{\rm vac} \left(\vec{E}\cdot \vec{A}^{(E)} - \vec{B}\cdot \vec{A}^{(B)}\right) \,,
\end{equation}

\ni a part of the interaction (13), collaborating in this case with the Standard-Model electromagnetic interaction $-e_f \bar{\psi}_f  \gamma^\mu \psi_f A^\mu$ of $f$ fermions. Then, in this channel, the total decay rates at rest, $\vec{k}_A = 0$, are ({\it cf.} Ref. [1]):

\begin{equation}
\Gamma\!\left(A^{ (E)} \rightarrow \bar{f} f\right) = \frac{e^2_f\, f \!<\!\!\varphi\!\!>^2_{\rm vac}}{24\pi M}\left[1 + \left(\frac{2m_f}{M}\right)^{\!\!2}\right] \left[1 - \left(\frac{2m_f}{M}\right)^{\!\!2}\right]^{\!1/2}  
\end{equation}

\ni and

\begin{equation}
\Gamma\!\left(A^{ (B)} \rightarrow \bar{f} f\right) = 0\,.
\end{equation}

When interacting in pairs, the sterile mediating $A$ bosons can annihilate simply into two photons, $A A \rightarrow  \varphi_{\rm ph}^*\! \gamma\, \varphi_{\rm ph}^*\!\gamma \rightarrow \gamma\gamma$, due to the coupling

\vspace{-0.3cm}

\begin{equation}
\sqrt{\!f\,}\,\varphi_{\rm ph} \left(\vec{E}\cdot \vec{A}^{(E)} - \vec{B}\cdot \vec{A}^{(B)}\right) 
\end{equation}

\ni being also a part of the interaction (13). In this channel, the S-matrix elements are (in the obvious notation):

\begin{eqnarray}
S\left(AA \rightarrow\!\!\!\!\right. & \!\!\left.\gamma \gamma\right) \!\! & \!\!= if \left[\frac{1}{(2\pi)^{12}}\frac{1}{16 \omega_1 \omega_2 \omega_{A1} \omega_{A2}}\right]^{1/2} (2\pi)^4 \delta^4(k_1 + k_2 - k_{A1} - k_{A_2})\times \nonumber \\ \nonumber \\
& & \!\times\left\{\left[ \begin{array}{c} (\omega_1 \vec{e}_1 - \vec{k}_1 e^0_1)\!\cdot\!\vec{e}^{\,(E1)\!}) \\ {\rm or} \\ - (\vec{k}_1 \times\!\vec{e}_1)\!\cdot\!\vec{e}^{\,(B1)\!}) \end{array}\right] \frac{1}{(k_{A1}  -  k_1\!)^2  -  m^2_\varphi} \left[ \begin{array}{c}(\omega_2\vec{e}_2  -  \vec{k}_2 e^0_2)\!\cdot\!\vec{e}^{\,(E2)}\!)\\ {\rm or} \\ - (\vec{k}_2\!\times\!\vec{e}_2)\!\cdot\!\vec{e}^{\,(B2)}\!)\end{array}\right] \right. \nonumber \\ \nonumber \\ 
& & \;\;\, +\left.\!\left[ \begin{array}{c} (\omega_2 \vec{e}_2 - \vec{k}_2 e^0_2)\!\cdot\!\vec{e}^{\,(E1)\!}) \\ {\rm or} \\ - (\vec{k}_2 \times\!\vec{e}_2)\!\cdot\!\vec{e}^{\,(B1)\!}) \end{array}\right] \frac{1}{(k_{A1}  -  k_2\!)^2  -  m^2_\varphi} \left[ \begin{array}{c}(\omega_1\vec{e}_1  -  \vec{k}_1 e^0_1)\!\cdot\!\vec{e}^{\,(E2)}\!)\\ {\rm or} \\ - (\vec{k}_1\!\times\!\vec{e}_1)\!\cdot\!\vec{e}^{\,(B2)}\!)\end{array}\right] \right\}, \nonumber \\ 
\end{eqnarray}

\ni where two identical photons $\gamma \gamma$ of momenta $\vec{k}_1, \vec{k}_2$ and linear polarizations $\vec{e}_1,\vec{e}_2$ are symmetrized (while $AA = A1A2$ with $A1 = E1$ or $B1$ and $A2 = E2$ or $B2$). The fully  differential cross-section is

\begin{equation}
\frac{d^{\,6}\sigma\!\left(AA \rightarrow \gamma \gamma\right)}{d^3\vec{k}_1 d^3\vec{k}_2} =  \frac{(2\pi)^3\!}{v_{\rm rel}} \sum_{e_1}\sum_{e_2}\frac{1}{3} \sum_{e^{(E1)}\,{\rm or}\,e^{(B1)}}\frac{1}{3} \sum_{e^{(E2)}\,{\rm or}\,e^{(B2)}} 
\frac{|S(AA \rightarrow \gamma\gamma)|^2}{(2\pi)^4 \delta^4(0)} \,,
\end{equation}

\ni while the total cross-section in this channel becomes

\begin{equation}
\sigma\!\left(\!AA\! \rightarrow \!\gamma \gamma\right) \!=\! \frac{1}{2} \int\! d^3\vec{k}_1 d^3\vec{k}_2 \frac{d^{\,6}\sigma\!\left(AA\! \rightarrow \!\gamma \gamma\right)}{d^3\vec{k}_1 d^3\vec{k}_2} \,.
\end{equation}

\ni We will use the electromagnetic gauge, where $e^0_1 = 0 = e^0_2$. In the centre-of-mass frame, $\vec{k}_{A1} + 
\vec{k}_{A2} = 0$, $\omega_{A1} = \omega_{A2}\;(\equiv \omega_{A})$ and $v_{\rm rel} = 2|\vec{k}_{A1}|/\omega_{A1} = 2v_{A1} = 2v_{A2} ( \equiv 2v_{A})$.

For the channel $A^{(E)}A^{(E)} \rightarrow \gamma \gamma$ (then, $\vec{e}^{\,(B1)} = 0 = \vec{e}^{\,(B2)}$ in Eq. (21)), we calculate from Eqs. (21), (22) and (23) in the centre-of-mass frame that

\begin{eqnarray}
\!\!\!\!\!\sigma\!\left(A^{(E)}\!A^{(E)} \!\!\rightarrow \!\gamma\gamma \right)\! 2v_A \!\!\!\!&\!=\!&\!\!\!\! \frac{1}{36} \frac{f^2}{(2\pi)^4}\! \left[\!\frac{1}{2m^2_\varphi\!+\! \frac{(M^2- m^2_\varphi)^2}{2\omega_A^2}} \right. \nonumber \\ 
\!\!\!\!& &\!\!\!\!\left.\!+ \frac{1}{4} \frac{1}{2\omega^2_A \!\left(1 \!-\! \frac{M^2- m^2_\varphi}{2\omega^2_A}\right)\!\sqrt{1\!-\! \frac{M^2}{\omega_A^2}}}\ln \!\left( 
\frac{ 1+ \sqrt{1- \frac{M^2}{\omega^2_A}} \!-\! \frac{M^2- m^2_\varphi}{2\omega_A^2} }{ 1\!-\!\sqrt{1\!-\!\frac{M^2}{\omega^2_A}} - \frac{M^2- m^2_\varphi}{2\omega_A^2} }\right)\!\right]\!\!, 
\end{eqnarray}

\ni where in the centre-of-mas frame it follows that $\vec{k}_1 + \vec{k}_2 = 0$, $\omega_1 = \omega_2 = \omega_{A1} =\omega_{A2}\;(\equiv \omega_{A})$ and $v_{A1} = v_{A2}\; (\equiv v_{A} = \sqrt{\omega^2_A - M^2}/\omega_A) $ as $|\vec{k}_{A1}| = |\vec{k}_{A2}| (\equiv \sqrt{\omega^2_A - M^2})$.

In the case of channel $A^{(B)}\!A^{(B)} \!\!\rightarrow \!\!\gamma\gamma $ (then, $\vec{e}^{\,(E1)} = 0 = \vec{e}^{\,(E2)}$ in Eq. (21)), we obtain the identical total cross-section (24) as in the case of $A^{(E)}\!A^{(E)} \!\!\rightarrow \!\!\gamma\gamma $, 

\begin{equation}
\sigma(A^{(B)}\!A^{(B)} \!\!\rightarrow \!\gamma\gamma)  = \sigma(A^{(E)}\!A^{(E)} \!\!\rightarrow \!\!\gamma\gamma )\,.
\end{equation}

\ni Finally, for the channel $A^{(E)}\!A^{(B)} \!\!\rightarrow \!\!\gamma\gamma $ (when $\vec{e}^{\,(B1)} = 0 = \vec{e}^{\,(E2)}$ {\it or} $\vec{e}^{\,(E1)} = 0 = \vec{e}^{\,(B2)}$), we get for $\sigma(A^{(E)}\!A^{(B)} \!\!\rightarrow \!\!\gamma\gamma)$ the formula differing from $\sigma(A^{(E)}\!A^{(E)} \!\!\rightarrow \!\!\gamma\gamma)$ given in Eq. (24) only by a minus sign at the front of logarithm term. Hence,

\begin{eqnarray}
\frac{1}{4}\left[\sigma\left(A^{(E)}A^{(E)} \rightarrow \gamma\gamma \right) + \sigma\left(A^{ (B)}A^{(B)} \rightarrow \gamma\gamma \right)\right.\!\! &\!\!+\!\!&\!\! \left.2\sigma\left(A^{(E)}A^{(B)} \rightarrow \gamma\gamma \right)\right]\;\;\;\;\;\;\;\;\;\;\;\;\;\;\;\;\;\;\;\;\;\; \nonumber \\
&\!\! = \!\! & \frac{1}{2v_A}\frac{1}{36} \frac{f^2}{(2\pi)^4} \frac{1}{2m^2_\varphi +\frac{(M^2-m^2_\varphi)^2}{2\omega^2_{A}}}\,.
\end{eqnarray}

\ni  If $f = e^2 = 4\pi \alpha$, then $(1/36) f^2/(2\pi)^4 = (1/36) \alpha^2/\pi^2\! = \!1.50\times 10^{-7}$ with $\alpha = 1/137$.

For production of sterile $A$ bosons, a simple channel is the inelastic Compton effect for $f$ fermions (where {\it e.g.} $f = e^-$ or $p$), $f \gamma \!\rightarrow\! f \gamma^* \!\rightarrow\! f A$, implied at the second step by $\gamma^* \!\rightarrow\! A$ with $\!<\!\!\varphi\!\!>_{\rm vac} \neq 0$, while for production of sterile $A$ boson pairs the simple channel $\gamma\gamma \!\rightarrow\! \varphi_{\rm ph}^*\!\gamma\,\varphi_{\rm ph}^*\!\gamma \!\rightarrow\! AA$ appears, inverse to the annihilation $AA \!\rightarrow\! \varphi_{\rm ph}^*\!\gamma\,\varphi_{\rm ph}^*\!\gamma \!\rightarrow\! \gamma\gamma$.

Dirac sterinos, candidates for cold dark matter [1], are stable in our model. In anti\-sterino-sterino pairs they can annihilate through the simple channels $\bar{\psi}\psi \!\!\rightarrow\!\! A^*\!\!\rightarrow \!\!\gamma^* \!\!\rightarrow \!\!\bar{f}f$ (with $<\!\!\varphi\!\!>_{\rm vac} \neq 0$ at the second step and $m_\psi > m_f$) and $\bar{\psi}\psi \!\rightarrow\! A^*\!\!\rightarrow \! \varphi_{\rm ph} \gamma$ (with $2m_\psi > m_\varphi$) having the total cross-sections in the centre-of-mass frame ({\it cf.} Ref. [1]): 

\begin{equation}
\sigma(\bar{\psi} \psi \rightarrow \bar{f} f) 2v_\psi = \frac{1}{12\pi}  \left(\frac{e\, f \zeta\!<\!\!\varphi\!\!>_{\rm vac}}{M^2}\right)^{\!\!2}\!\!\left(1+\frac{2m^2_\psi}{E^2_\psi}\right)\! \left(1+\frac{m^2_f}{2E^2_\psi}\right)\! \left(1-\frac{m^2_f}{E^2_\psi}\right)^{\!\!1/2} 
\end{equation}

\ni  and

\begin{equation}
\sigma(\bar{\psi} \psi \rightarrow \varphi_{\rm ph} \gamma) 2v_\psi = \frac{1}{6\pi} 
\left(\frac{f \zeta}{M^2}\right)^{\!\!2} \left(1+ \frac{2m^2_\psi}{E^2_\psi}\right) \left(E^2_\psi -   \frac{m^2_\varphi}{4}\right) \,.
\end{equation}

Physical sterons are unstable, decaying in the simple channel $\varphi_{\rm ph} \!\rightarrow\! A^*\gamma \!\!\rightarrow\! \gamma\gamma$ (where $<\!\!\varphi\!\!>_{\rm vac} \neq 0$ at the second step) with the total rate at rest ({\it cf.} the second Ref. [2]): 

\begin{equation}
\Gamma(\varphi_{\rm ph}\!\rightarrow\! \gamma\gamma) = \frac{1}{128\pi} \!\left(\frac{f<\!\!\varphi\!\!>_{\!\rm vac}}{M^2} \right)^{\!\!2} \,m^3_\varphi \;. 
\end{equation}

\ni  If $f = e^2 = 4\pi \alpha$, then $ f^2/(128\pi) = \alpha^2\pi/8 = 2.09\times 10^{-5}$ with $\alpha = 1/137$.

We obtain the results (27), (28) and (29) applying for simplicity the effective interaction (5) (with the approximately eliminated $A_{\mu \nu}$) to the processes $\bar{\psi} \psi \rightarrow \gamma^* \rightarrow  \bar{f}f$ (where $<\!\!\varphi\!\!>_{\rm vac} \neq 0$ at the first step), $\bar{\psi} \psi \rightarrow \varphi_{\rm ph} \gamma$ and $\varphi_{\rm ph} \rightarrow \gamma\gamma$ (where $<\!\!\varphi\!\!>_{\rm vac} \neq 0$ also).

\vspace{0.3cm}

\ni {\bf 4. Conclusions and final remarks} 

\vspace{0.3cm}

In the presented model of hidden sector of the Universe there are three new proposals being to some extent unusual:

\vspace{0.2cm}

\ni (i) Interactions within the hidden sector are mediated by a nongauge antisymmetric-tensor field $A_{\mu \nu}$ (of dimension one). It means that gauge interactions, defining some charges, are restricted to appear only in the \SMo sector.

\ni (ii) Photons, described in a gauge-invariant way by the electromagnetic field $F_{\mu \nu} = \partial_\mu A_\nu - \partial_\nu A_\mu$ (after the electroweak symmetry is spontaneously broken by the \SMo Higgs mechanism), participate both in the familiar \SMo electromagnetic coupling $-j^\mu A_\mu$ to the electric current $j^\mu$ and in a new hidden-sector weak coupling $-(1/2)\sqrt{f\,} \varphi F_{\mu \nu} A^{\mu \nu}$ to the product field $\varphi A^{\mu \nu}$, where $\varphi$ is a sterile spin-0 field ("$\!$\,steron"\, field) required to exist in order to form together with the product field $F_{\mu \nu} A^{\mu \nu}$ the new Lorentz- and gauge-invariant coupling (of dimension four).

\ni (iii) The field $\varphi$ gets spontaneously a nonzero vacuum expectation value $<\!\!\varphi\!\!>_{\rm vac} \neq 0$, defining the physical field  $\varphi_{\rm ph} = \varphi - <\!\!\varphi\!\!>_{\rm vac}$ .

\vspace{0.15cm}

In view of the above conjectures it is natural to introduce also a sterile spin-1/2 Dirac field $\psi$ ("$\!$\,sterino"\, field), interacting in the hidden sector through the weak coupling $-(1/2)\sqrt{f\,}\zeta \bar{\psi} \sigma_{\mu \nu} \psi A^{\mu \nu}$.

Of course, the universal gravity ought to be active for all hidden-sector particles as it is for all \SMo particles.

The above proposals suggest in particular the following features of our model:

\vspace{0.2cm}

\ni (1) The hidden and  \SMo sectors interact weakly due to photons coupled in both sectors ("photonic portal"\, to the hidden sector). Effectively, this weak interaction gets a magnetic character, though the hidden sector is electrically neutral. It follows from the form $j^\mu + \sqrt{f\,} \partial_\nu(\varphi A^{\mu \nu})$ of the total source current for $F^{\mu \nu}$ in the "$\!\,$supplemented Maxwell's equations", where the four-divergence describes the contribution from hidden sector.

\ni (2) Sterile mediating $A$ bosons are unstable. Due to $<\!\!\varphi\!\!>_{\rm vac} \neq 0$, photons are (linearly) a part of their source.

\ni (3) Dirac sterinos are stable candidates for cold dark matter.

\ni (4) They display small spontaneously generated magnetic moment, though they are electrically neutral.

\ni (5) Sterons are unstable.

\vspace{0.2cm}

Note that the structure of the Universe proposed here can be embedded in a more extended one, displaying the overall electroweak symmetry spontaneously broken by the \SMo Higgs mechanism ({\it cf.} the third Ref. [2]).

Obviously, a careful study of experimental consequences of the proposed model is a task that ought to be undertaken in order to justify our proposals. I hope that such a work may turn out to be attractive from the experimental point of view.

As far as matter fermions are concerned, the proposed model of the Universe has a two-level pyramidal structure 

\vspace{0.2cm}

{\centerline{sterinos}}

{\centerline{leptons and quarks}}

\ni with

$$
{\rm leptons} = \left\{\begin{array}{ll} e^-\, , & \!\!\nu_e \\ \mu^- ,& \!\!\nu_\mu \\ \tau^- ,& \!\!\nu_\tau \end{array}\right.\;\;,\;\; 
{\rm quarks} = \left\{\begin{array}{ll} u\; ,& \!\!d \\ c\; ,& \!\!s \\ t\; ,& \!\!b \end{array}\right.\;{\rm (in\;three\;colors})\;,
$$
 
\vspace{0.3cm}

\ni where the sterino level is built up formally by removing from fundamental leptons and quarks their \SMo charges: electric and weak for leptons, and electric, weak and color for quarks. Such an operation may be performed in reality by an intrinsic formalism of generalized Dirac equations, following generically from the familiar Dirac square-root procedure as it has been invented several years ago [5,6]. In this formalism, all sterinos and all leptons and quarks can be described by the master wave functions $\psi_{\beta_1...\beta_N}(x)\,(N = 1,3,...)$ and $\psi_{\alpha \beta_1...\beta_N}(x)\,(N = 0,2,4,...)$, respectively, satisfying the gene\-ralized Dirac equations. These functions carry multiple Dirac bispinor indices, namely $\beta_1,...,\beta_N$ and $\alpha,\beta_1,...,\beta_N$, respectively, where $\beta_1,...,\beta_N$ are assumed to be physically undistinguishable and chosen antisymmetrized ("intrinsic Pauli principle"), while $\alpha$ is physically distinguished from $\beta_1,...,\beta_N$, when it is presumed to be accompanied by the \SMo $SU(3)\!\times\! SU(2)\!\times\! U(1)$ labels of leptons and quarks (suppressed in our notation). Then, in this approach there appear two and only two generations $N = 1,3$ of sterinos,

\begin{equation}
{\rm sterinos} = \left\{\begin{array}{lcl} \psi^{(1)}_\beta (x) & = & \psi_\beta (x)\,,  \\ \psi^{(3)}_\beta (x) & =  & \frac{1}{6} \varepsilon_{\beta_ 1\beta_ 2\beta_ 3\beta}\psi_{\beta_ 1\beta_ 2\beta_ 3} (x)\,,\end{array} \right. 
\end{equation}

\ni and three and only three generations $N = 0,2,4$ for leptons and quarks,

\begin{equation}
{\rm leptons\;and\;quarks} = \left\{\begin{array}{lcl} \psi^{(0)}_\alpha (x) & = & \psi_\alpha (x)\,,  \\ \psi^{(2)}_\alpha (x) & = & \frac{1}{4}(C^{-1}\gamma_5)_{\beta_ 1\beta_ 2}\psi_{\alpha \beta_1\beta_2}(x)\,, \\ \psi^{(4)}_\alpha (x) & = & \frac{1}{24}\varepsilon_{\beta_1\beta_2\beta_3\beta_4}\psi_{\alpha \beta_1\beta_2\beta_3\beta_4}(x)\,.  
\end{array}\right. 
\end{equation}

\vspace{0.1cm}

\ni The latter conclusion justifies three experimental generations for leptons and quarks as a consequence of the familiar Dirac square-root procedure (realized generically) and the new "intrinsic Pauli principle".\footnote{One may say that the elementary matter fermions, sterinos or leptons and quarks, are here local objects composed of spin-1/2 "intrinsic partons"\, characterized by the Dirac bispinor indices $\beta_1,...,\beta_N\;(N = 1,3)$ or $\alpha, \beta_1,...,\beta_N\;(N = 0,2,4)$, respectively, and, in the case of "intrinsic parton"\, with the index $\alpha$, additionally by the $SU(3)\times SU(2)\times U(1)$ labels (in a suppressed way). The "intrinsic partons"\, with indices $\beta_1,...,\beta_N$ are undistinguishable and obey Fermi statistics along with the "intrinsic Pauli principle".} 

In the first Ref. [6] there are mentioned some possible formal objections against two sterino generations, suggesting instead the existence of three such generations, similarly as for leptons and quarks. The argument is that among $N$ sterino Dirac indices $\beta_1,...,\beta_N$ one might be physically distinguished by our nongauge hidden-sector interactions, leaving only $N-1$ of them antisymmetrized. Then, denoting this distinguished Dirac index by $\beta$ and the antisymmetrized rest by $\beta_2,...,\beta_N$, we might have the master wave function $\psi_{\beta\beta_2...\beta_N}(x)$ and so, three generations $N-1 = 0,2,4$ for sterinos,

\begin{equation}
{\rm sterinos} = \left\{\begin{array}{lcl} \psi^{(0)}_\beta (x) & = & \psi_\beta (x)\,,  \\ \psi^{(2)}_\beta (x) & = & \frac{1}{4}(C^{-1}\gamma_5)_{\beta_2\beta_3}\psi_{\beta \beta_2\beta_3}(x)\,, \\ \psi^{(4)}_\beta (x) & = & \frac{1}{24}\varepsilon_{\beta_2\beta_3\beta_4\beta_5}\psi_{\beta \beta_ 2\beta_ 3\beta_4\beta_5}(x)\,.  
\end{array}\right. 
\end{equation}

\ni Of course, no $SU(3)\!\times\!SU(2)\!\times\!U(1)$ labels accompany here (in a suppressed way) the Dirac index $\beta$, although it is physically distinguished from the antisymmetrized $\beta_2,...,\beta_N$.{\footnote{Then, in sterinos, the "intrinsic parton"\, with the Dirac bispinor index $\beta$ is characterized additionally (in a suppressed way) by its ability to take part in nongauge hidden-sector interactions.}}

\vfill\eject

\vspace{0.4cm}

{\centerline{\bf References}}

\vspace{0.4cm}

\baselineskip 0.73cm

{\everypar={\hangindent=0.65truecm}
\parindent=0pt\frenchspacing

{\everypar={\hangindent=0.65truecm}
\parindent=0pt\frenchspacing

[1]~W.~Kr\'{o}likowski, arXiv: 0909.2498 [{\tt hep--ph}]. 

\vspace{0.2cm}

[2]~W.~Kr\'{o}likowski, {\it Acta Phys. Polon.} {\bf B 39}, 1881 (2008); {\bf B 40}, 111 (2009); {\bf B 40}, 2767 (2009).

\vspace{0.2cm}

[3]~{\it Cf. e.g.} M.~Ahlers, L.A.~Anchordoqui and M.C.~Gonzalez-Garcia , arXiv: 0910.5483 [{\tt hep-ph}]. 

\vspace{0.2cm}

[4]~{\it Cf. e.g.}  J. March-Russell, S.M. West, D. Cumberbath and D.~Hooper, {\it J. High Energy Phys.} {\bf 0807}, 058 (2008); K.~Kohri, J.~McDonald and N.~Sahu, arXiv: 0905.1312 [{\tt hep-ph}]; and references therein.

\vspace{0.2cm}

[5]~W.~Kr\'{o}likowski, {\it Phys. Rev.} {\bf D 45}, 3222 (1992); {\it Acta Phys. Polon.} {\bf B 33}, 2559 (2002); {\tt hep--ph}/0504256; {\it Acta Phys. Polon.} {\bf B 38}, 3133 (2007); and references therein.

\vspace{0.2cm}

[6]~W.~Kr\'{o}likowski, arXiv: 0811.3844 [{\tt hep--ph}]; arXiv: 0812.1875 [{\tt hep--ph}].

\vspace{0.2cm}

\vfill\eject

\end{document}